\documentclass[letterpaper,aps,prl,twocolumn,amsmath,showpacs,amssymb]{revtex4}

\usepackage{color}
\pdfoutput=1
\usepackage[pdftex]{graphicx,hyperref}
\usepackage{dcolumn}
\usepackage{bm}
\usepackage{bbm}
\usepackage{wasysym}
\usepackage{marvosym}
\usepackage{times,amsmath,amssymb}

\def\lapp{\lower.35em\hbox{$\stackrel{\textstyle<}{\sim}$}}

\newlength{\textwidthm}

\begin{document}

\title{Excitonic effects in the optical conductivity of gated graphene}

\author{N. M. R. Peres$^1$, R. M. Ribeiro$^1$, and A. H. Castro Neto$^2$} 

\affiliation{$^1$ Department of Physics and Center of Physics,
 University of  Minho, P-4710-057, Braga, Portugal}

\affiliation{$^2$
Department of Physics, Boston University, 
590 Commonwealth Avenue, Boston, Massachusetts 02215, USA}

\date{\today}

\begin{abstract}
We study the effect of electron-electron interactions in the optical 
conductivity of graphene under applied gate and derive a generalization of Elliott's 
formula, commonly used for semiconductors, for the optical intensity. 
We show that {\it excitonic resonances} are responsible for several 
features of the experimentally measured mid-infrared response of graphene 
such as the increase of the conductivity beyond the ``universal'' value above 
the Fermi blocked regime, the broadening of the absorption at the threshold, 
and the decrease of the optical conductivity at higher frequencies. 
\end{abstract}

\pacs{81.05.ue,72.80.Vp,78.67.Wj}

\maketitle

Since its isolation in 2004 \cite{nov04}, the demonstration of 
the presence of electron-electron interactions in the electronic 
properties of graphene has been elusive \cite{rmp}. The first clear
 manifestation of interaction effects comes from the recent measurement 
of the fractional quantum Hall effect at filling $1/3$ \cite{Evafrac,Kimfrac}. 
To understand the impact of many-body physics in graphene and disentangle it 
from purely band structure effects it is essential to fully understand graphene's electronic
and transport properties. 
Furthermore, because Coulomb interactions can be controlled externally by choosing 
the dielectric constant of the substrate on which graphene is deposited, it is
 possible to tune the electron-electron interactions in a way that cannot be done in most materials.

An obvious property where to look for correlation effects is the optical 
response. It is well established that in ordinary semiconductors 
electron-electron interactions are essential to  explain their 
optical properties \cite{Haug2}. For energies below the gap, a 
particle and a hole created by the excitation process can form a 
bound state (an exciton), which appears as a well defined peak in 
the absorption spectra, in the region within the energy gap. 
For energies above the gap, a renormalization of the band 
and an enhancement of the optical conductivity is observed 
and attributed to the Coulomb-mediated scattering between the 
electron-hole pair. These effects are embodied in the 
famous Elliott's formula \cite{Haug2} that describes the intensity 
of optical absorption close to the gap edge. This is true for 
traditional semiconductors in both three and two dimensions \cite{Haug1}. 
As we are going to show the situation in graphene is, however, quite different.

Firstly, in graphene there is no gap between the 
lower ($\pi$) and upper ($\pi^*$) bands, the effective theory describing graphene's 
low energy physics is the massless two-dimensional Dirac equation, and the Coulomb 
problem in this case has no true bound states, but resonances 
\cite{lin,coulombpereira,coulombnovikov,coulomblevitov}. These 
facts prevent the formation of bound excitons in graphene. 
On the other hand, {\it exciton resonances} can exist and a 
signature of their presence should be seen in its optical 
properties \cite{kuzmenko,nair,mak,basov,Crommieopt}. Although the system 
is gapless, the possibility of  doping it by using a gate creates 
an energy interval, of width $2\mu$  ($\mu$ is chemical potential), 
where light absorption is forbidden due to Pauli's exclusion principle. 
In tradicional semiconductors,
such a regime is known as the Moss-Burshtein effect. 
All theoretical investigations of the optical conductivity of gated 
graphene developed so far evaded the problem of electron-electron 
interactions and the discussion of excitonic effects 
\cite{nmrPRB06,falkovsky,stauberphonons,carbotte,Juan}. 
However, a detailed analysis of the experimental data \cite{nair,basov} 
shows that graphene's optical conductivity deviates considerably from the 
non-interacting prediction \cite{nair,basov}, even with the inclusion of 
band structure corrections to the Dirac spectrum \cite{staubergeim}, 
disorder, and phonons \cite{stauberphonons}. 

\begin{figure}
\includegraphics*[width=8.5cm]{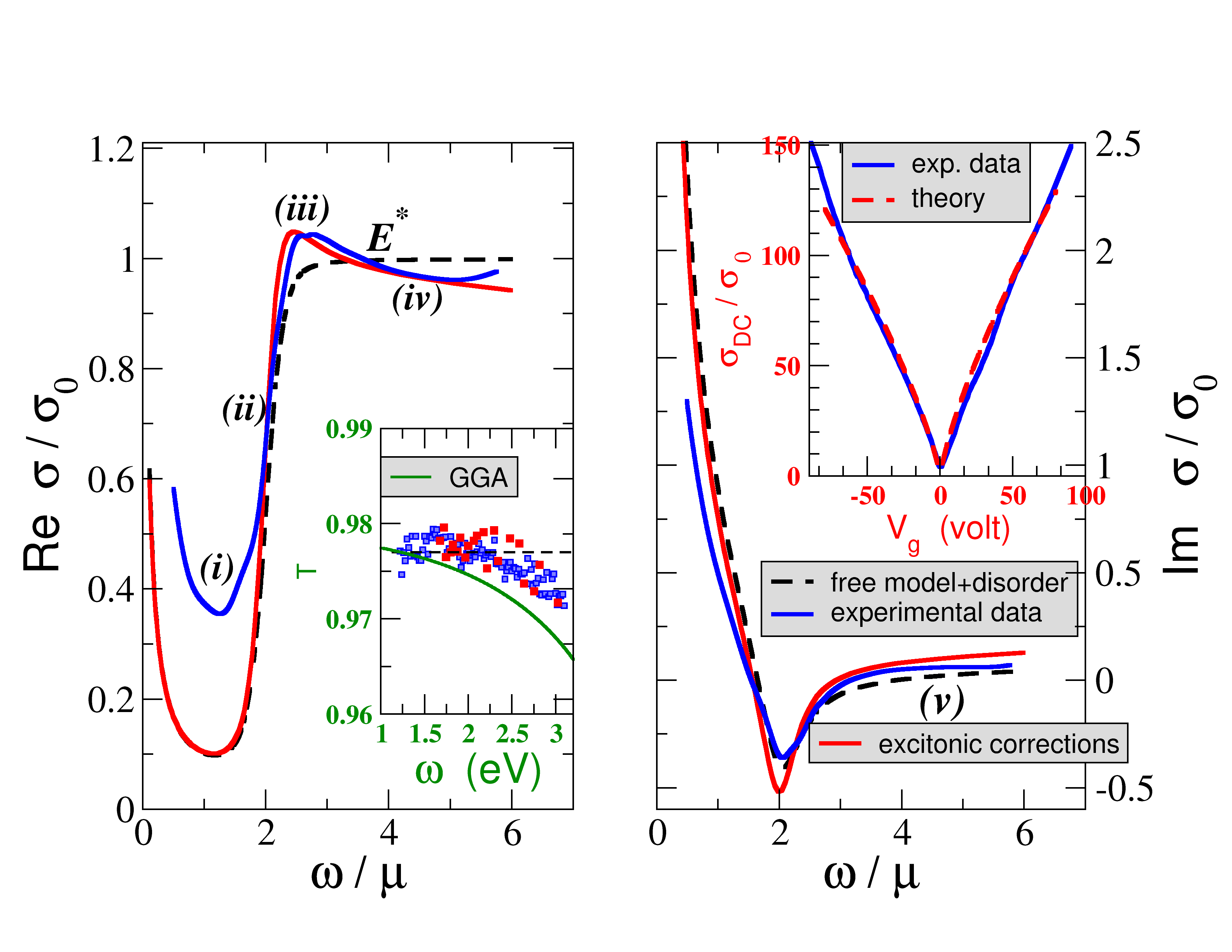}
\caption{(Color online) Real and imaginary parts of the conductivity.
The black dashed line is obtained from the independent electron model
with disorder. The solid red line is the model given in the text,
which includes excitonic corrections and disorder. The experimental curves, refer to
a gate voltage of 28 V, to which corresponds a Fermi energy of $\mu\simeq0.18$ eV.
In the experiment\cite{basov}, the chemical potential was 
given by $\mu=0.03474\sqrt V_g$ eV. 
The calculation took an effective temperature of
120 K (see the discussion in point
({\it ii}) in the text).
In the left inset we give the transmission, $T$, of
neutral graphene, as function of $\hbar\omega$, in comparison with the 
experimental data of Ref. \onlinecite{nair}. 
In the right inset we show the DC conductivity as a function of 
gate voltage,  comparing both the experimental data (blue line) 
of Ref. \onlinecite{basov} and
the thoretical fit;
the fit fixes the concentration of impurities, 
to the values of $n_i=2\times 10^{11}$ cm$^{-2}$, for resonant
scatterers, and $n_i=1\times 10^{11}$ cm$^{-2}$, for charged ones.
The determined impurity concentrations are then used in the 
calculation of the optical conductivity. Therefore, the calculation of the
optical conductivity has no fitting parameter. 
The labels (i), (ii), (iii), (iv), and
(v), refer to the five items discussed in the text.}
\label{fig:optical_conductivity}
\end{figure}

The alluded deviations occur 
in both neutral and doped graphene, but they are far more evident in the latter case. 
For neutral graphene, in the visible range of the spectrum at energies 
around $\sim 3$ eV, the measured percentage of transmitted light, $T$, 
is smaller than the value $T=1-\pi\alpha\simeq 97.7\%$ ($\alpha$ is 
the fine structure constant), predicted by the non-interacting theory \cite{nair}. 
Band structure effects alone cannot account for the measured deviations \cite{staubergeim}. 
In the left inset of Fig. \ref{fig:optical_conductivity} it is shown that 
calculations based on {\it ab-initio} methods can describe,
qualitatively, the data (see also Ref. \onlinecite{LiLouie}).

For gated graphene, experiments \cite{basov}  show even more noticeable
deviations from the non-interacting picture, which predicts 
that the optical conductivity of graphene has the form: 
$\sigma(\omega)=\sigma_0\theta(\hbar\omega-2\mu)$, 
where $\sigma_0=\pi e^2/(2h)$ is the so-called ``universal'' 
AC-conductivity, $\hbar\omega$ is the photon energy, and $\theta(x)$ 
is the Heaviside step function. The deviations seen in the data 
are of five different types: ({\it i}) finite absorption below $2\mu$, 
which is due to both inter-band and intra-band elastic and inelastic 
scattering processes; ({\it ii}) broadening of the absorption edge 
around the energy threshold $2\mu$; ({\it iii}) an enhancement of the 
conductivity above the universal value, $\sigma_0$, in the energy 
range between $2\mu$ and $2\mu+E^\ast$, where $E^\ast$ is a 
characteristic energy scale; ({\it iv}) a reduction of the 
conductivity bellow $\sigma_0$, at energies above $E^\ast$, 
with the conductivity, as a function of frequency, having a 
positive curvature; ({\it v}) the imaginary part of the 
conductivity is 
larger than the value predicted by the non-interacting model 
for energies $\hbar\omega\gg2\mu$. Finally, we point out that 
the optical conductivity curves, for diferent gate voltages
\cite{basov}, collapse on top of each other, 
when re-plotted as function of $\omega/\mu$, implying that the mechanism causing 
deviations from the non-interacting approximation must be intrinsic. 
The analytical microscopic theory we develop in this letter, which includes excitonic effects,
 accounts for items ({\it ii}), ({\it iii}), ({\it iv}), and ({\it v}),
 and it also partially accounts for item ({\it i}), although not completely, 
since we have not included intra-band scattering in the calculations.

The central result of this work is the calculation 
of the conductivity of doped graphene, that is, at 
finite gate voltage, $V_g$. The calculation takes into 
account electron-electron interactions at the exchange level. 
Its important to stress that the theory given below
has no adjustable parameter (except for the use of an effective temperature --see discussion below);
the density of impurities is determined by fitting the DC conductivity, that
is, it is  fixed {\it a priori}
to the calculations of the optical conductivity.
The final result for the latter quantity reads: 
\begin{equation}
\sigma=-\frac{ev_F}{\pi^2}\int d\,\bm k\left( \Gamma_{vc,\bm k}\chi^0_{vc,\bm k}d^x_{vc,\bm k}+
\Gamma_{cv,\bm k}\chi^0_{cv,\bm k}d^x_{cv,\bm k}\right)\,,
\label{eq:sigmaxx}
\end{equation}
where all quantities, to be given below, 
depend explicitly on $\omega$, and on temperature. 
In Eq. (\ref{eq:sigmaxx}),  $e>0$ is the electron charge, $v_F$ 
is the Fermi velocity, $d^x_{\lambda\lambda',\bm k}$ is the matrix 
element of the dipole operator with $\lambda,\lambda'=c,v$ 
referring to the $\pi^*$ ($\pi$) band index, respectively, 
$\chi^0_{\lambda\lambda',\bm k}$ is the optical susceptibility of 
graphene in the independent electron approximation, 
and $\Gamma_{\lambda\lambda',\bm k}$ is the vertex 
correction due electron-electron interactions. If we 
assume $\Gamma_{\lambda\lambda',\bm k}=1$, Eq. (\ref{eq:sigmaxx}) 
gives the ``universal'' value for the conductivity of graphene, $\sigma_0$. 
The form of the vertex function reads: 
$\Gamma_{\lambda\lambda',\bm k}=[1+s_{vc,\bm k,1}]^{-1}$, 
where $s_{vc,\bm k,1}$ is defined below. This result 
shows that if $\Gamma_{\lambda\lambda',\bm k}<1$, its 
contribution to Eq. (\ref{eq:sigmaxx}) acts to 
reduce $\sigma$ relatively to $\sigma_0$. On the 
other hand, when $\Gamma_{\lambda\lambda',\bm k}>1$ it 
contributes to the enhancement of $\sigma$ over $\sigma_0$.
 We show below that the effect of the vertex correction on $\sigma$ depends on $\omega$.  
The conclusion is then:  the importance of electron-electron interactions 
in graphene can manifest itself in the behavior of $\sigma$ by studying how 
the data deviates from $\sigma_0$. Moreover, the optical conductivity relates 
to the transmittance by: $T=[(1+\Re\sigma/(2\epsilon_0c))^2+(\Im\sigma/(2\epsilon_0c))^2]^{-1}$, 
where $\epsilon_0$ is the vacuum permittivity and $c$ is the speed of light. 
Our results are summarized in Fig. \ref{fig:optical_conductivity}. In the 
left panel, we depict the real part of the conductivity: the dashed black 
curve is the non-interacting prediction, considering the effect of disorder; 
the solid red curve is the prediction taking into account excitonic effects 
and the same amount of disorder. The right panel refers to the imaginary 
part of the conductivity.
In both panels the experimental data \cite{basov} is also shown.

The five items mentioned previously relatively to the data,
can  easily be observed to be present in the 
theoretical curve; ({\it i}): For $\hbar\omega<2\mu$, 
there is a finite conductivity, where the non-disordered independent electron 
model predicts zero absorption. This modification over the naive result is mainly a 
consequence of disorder, and excitonic effects play no role, as it should be.
The discrepancy of about a factor of two between the theory and the data, is to be
expected and is a strength of the model, since 
 additional
intra-band scattering (the model only includes inter-band scattering) 
will transfer spectral weight from the Drude peak 
to this frequency region, enhancing the optical response.
({\it ii}): 
It is clear that disorder and temperature effects combined, account for 
the broadening of the Fermi step around $2\mu$. As discussed in the context
of the optical response of the biased graphene bilayer \cite{kuzmenko2}, 
charge density fluctuations,
associated with the random electrostatic potential present at the graphene-SiO$_2$ interface,
smear the Fermi energy, leading to 
and effective higher temperature than that measured by the finger temperature of
the cryostat
(in the experimente \cite{basov}, the temperature of the finger was
54 K); as it is well known, an increase of the temperature broadens
Fermi edge \cite{staubergeim}, as seen in the experiments (we have used the same 
effective temperature as that used in bilayer graphene, 120 K; this makes sense
because the random electrostatic potential alluded above is a property of the substrate). 
({\it iii}): For $\hbar\omega>2\mu$, an enhancement of the 
conductivity above $\sigma_0$ is seen. This behavior extends to the characteristic 
energy scale $E^\ast$. The energy interval over which $\Re\sigma>\sigma_0$ is 
about the same in the theory and in the data.  
Finally, we should stress that the results 
of Fig. \ref{fig:optical_conductivity} were obtained from an approximate 
analytical solution to the vertex functions. Corrections to this approximation 
may also account for the small differences between the data and the 
theoretical curves.
({\it iv}): For $\hbar\omega>E^\ast$, the conductivity 
goes below $\sigma_0$, over a large energy range, presenting a positive curvature.
 Exactly the same behavior is seen in our calculations, with the decrease of the 
conductivity below $\sigma_0$ having about the same numerical value. ({\it v}): The imaginary part of the 
conductivity, $\Im{\sigma}$, is larger, for energies above $\sim 3\mu$, 
than that predicted by the 
non-interacting model with disorder  and 
fits well the data.
In conclusion, Eq. (\ref{eq:sigmaxx}) 
does explain all features, {\it (ii)-(iv)}, of the experimental data \cite{basov},
making clear that electron-electron interaction effects are manifestly present in the
optical conductivity of gated graphene for $\hbar\omega\ge 2\mu$.
Finally, we must note in passing that we have chosen to fit a set of data where
no renormalization of the Fermi velocity was measured \cite{basov}, since our
 model does not include the renormalization of the electronic velocity 
For all data sets such that $V_g\ge 28$ V, no Fermi velocity renormalization was measured; we have chosen
the first of this class of sets, namely that for which $V_g=28$ V.

Before discussing the calculation of the electron-electron corrections to the optical conductivity, 
it is instructive to show that the formalism based on the polarization concept gives the 
well known result for the ``universal'' conductivity of graphene. As it will be clear, 
this formalism  bypasses the calculation of the Kubo formula \cite{stauberphonons,carbotte}. 
The low-energy Hamiltonian of graphene, with a vector potential, $\bm A(\bm r,t)$, reads:
$
H=v_F\bm \sigma\cdot [\bm p + e\bm A(\bm r,t)]=H_0+H_I\,. 
$
Assuming an electric field of the form $\bm E(t)=E_0 {\hat{\bf x}} e^{-i\omega t}$, the interaction 
of the electrons with light is given by  $H_I=ev_FA_0\sigma_xe^{-i\omega t}=ev_F\sigma_x A_0(t)$, 
with $A_0=-iE_0/\omega$. The polarization along the $x-$direction is:
\begin{equation}
P_x(t)=\sum_\sigma \int d\,\bm r
\langle
\hat\psi^\dag_\sigma(\bm r,t)d_x
\hat\psi_\sigma(\bm r,t)
\rangle\,,
\label{eq:polarization_field}
\end{equation}
where $d_x$ is the dipole operator, $d_x=-v_Fe\sigma_x$, and $\hat\psi_\sigma(\bm r,t)$ is
 the field operator:
\begin{equation}
 \hat\psi_\sigma(\bm r,t)=\sum_{\bm k,\lambda}
a_{\lambda,\bm k,\sigma}(t)\psi_{\lambda,\bm k}(\bm r)\,,
\label{eq:field_operator}
\end{equation}
with $\psi_{\lambda,\bm k}(\bm r)$ the eigenfunctions of $H_0$ 
\cite{rmp}, $a_{\lambda,\bm k,\sigma}$ the destruction operator of an 
electron in band $\lambda$, with momentum $\bm k=(k_x,k_y)$, and spin projection $\sigma$ 
($\lambda=1$ for $\lambda=c$ and $\lambda=-1$ for $\lambda=v$).  The matrix 
element of the dipole operator reads:
$
 d_{\lambda,\lambda',\bm k}^x=(\lambda'e^{i\theta(\bm k)}
+\lambda e^{-i\theta(\bm k)})/2$,
with $\tan\theta(\bm k)=k_y/k_x$. Then, the final expression for the 
polarization (per unit time) $P_x(t)$ is, in second quantized form, given by:
\begin{equation}
 P_x(t)=-Ev_F\sum_{\sigma,\lambda,\lambda'}\sum_{\bm k}
\langle  a^\dag_{\lambda,\bm k,\sigma}(t)a_{\lambda',\bm k,\sigma}(t)
\rangle d_{\lambda,\lambda',\bm k}^x\,,
\label{eq:polarization_momentum_final}
\end{equation}
and $H_I$ reads:\\
$
 H_I=v_FeA(t)
\sum_{\sigma,\lambda,\lambda'}\sum_{\bm k}
a^\dag_{\lambda,\bm k,\sigma}a_{\lambda',\bm k,\sigma}
 d_{\lambda,\lambda',\bm k}^x\,.
$

We introduce the dipolar operator $\hat P_{vc,\bm k}=a^\dag_{v,\bm k}a_{c,\bm k}$,
and seek  the  solution of  its 
equation of motion,
$-i\hbar\partial \hat P_{vc,\bm k}/\partial t=[H,\hat P_{vc,\bm k}]$. The procedure is 
simple, and gives the following result for the thermal average of the operator $\hat P_{vc,\bm k}$:
\begin{equation}
\langle \hat P_{vc,\bm k}\rangle= 
v_FeA_0d^x_{cv,\bm k}\frac{n_F(E_{c,\bm k})-n_F(E_{v,\bm k})}{-\omega\hbar-i\Gamma-
E_{v,\bm k}+E_{c,\bm k}}\,,
\end{equation}
where $n_F(x)$ is the Fermi function, $E_{c/v,\bm k}=\pm v_F\hbar k$, 
and $\Gamma(\omega)$ is the level broadening (a similar equation follows 
for $\langle \hat P_{cv,\bm k}\rangle$). We note that the broadening $\Gamma(\omega)$ is 
not a constant, but follows a well prescribed model, where the scattering is dominated 
by resonant and charged scatterers  \cite{stauberBZ,coulombnovikov}: 
$\Gamma(\omega)=n^{(\rm rs)}_iv_F\hbar\sigma^{(\rm rs)}_T(\omega)+n^{(\rm ch)}_iv_F\hbar\sigma^{(\rm ch)}_T(\omega)$, 
where $n_i^{(\rm rs/ch)}$ is the concentration of 
impurities and $\sigma_T^{(rs/ch)}(\omega)$ is the exact transport cross section, for each type
of scatterers. 
Once $\langle \hat P_{\lambda \lambda',\bm k}\rangle$ is known, the calculation of the
 polarization per unit area, $p_x=P_x(\omega)/A$, follows from
 Eq. (\ref{eq:polarization_momentum_final}) as:
\begin{equation}
p_x  = -i\frac{2e^2v_F^2E_0}{\omega A}\sum_{\bm k,
\lambda\ne\lambda '}s(\bm k)
\frac{n_F(E_{\lambda,\bm k})-n_F(E_{\lambda',\bm k})}
{\omega\hbar-E_{\lambda,\bm k}+E_{\lambda',\bm k}+i\Gamma}\,,
\label{eq:polarizatio_per_area}
\end{equation}
where $s(\bm k)=\sin^2[\theta(\bm k)]$, and the optical conductivity 
is simply given by $\sigma=p_x/E_0$. Equation (\ref{eq:polarizatio_per_area}) 
for the conductivity of graphene is the same one obtained by computing the Kubo 
formula from the current-current correlation function \cite{staubergeim}. If we
 consider, as a simple example, the limiting case where $\Gamma\rightarrow 0$, $\mu=0$, and 
zero temperature, Eq. (\ref{eq:polarizatio_per_area}) gives $p_x/E_0= \pi e^2/(2 h)=\sigma_0$.

We can extend the formalism above to the case where electron-electron interactions 
are considered, and use it to derive the central result of this work, Eq. (\ref{eq:sigmaxx}). 
The formalism we present below, gives results equivalent to the solution of the
Bethe-Salpeter equation, but following a more straightforward path.
Electron-electron interactions in graphene are described by the Hamiltonian:
\begin{equation}
H_{e-e}=\frac 1 2\int d\,\bm r d\,\bm r'
\hat \psi^\dag(\bm r) \psi^\dag(\bm r')V(\bm r-\bm r') 
\psi(\bm r') \psi(\bm r)\,,
\label{eq:Hee_space}
\end{equation}
where $V(\bm r)$ is the Coulomb potential. Since we are studying the 
conductivity of gated graphene, the Coulomb potential is screened, and at the 
level of the Thomas-Fermi approximation (corresponding to the static RPA limit) 
the Fourier transform of the Coulomb potential reads: 
$
\varphi_{\rm 2D}(\bm q) =e^2 
[2\epsilon_0(q+q_{\rm TF})]^{-1}\,,
$
with the Thomas-Fermi momentum, $q_{\rm TF}$, defined 
as $q_{\rm TF}=4\alpha_g\sqrt{\pi n_e}$, where $n_e$ is the electron density 
per unit area and $\alpha_g=e^2/(4\pi\epsilon_0v_F\hbar)$ (for graphene 
on top of SiO$_2$ we replace $\epsilon_0$ by $\epsilon_0 \epsilon_{\rm SiO_2}$ 
and $e^2/(4\epsilon_0\epsilon_{\rm SiO_2}v_F\pi\hbar)\approx 1/2$;
in our calculations we used $\alpha_g=0.4$).  
Introducing the field operators (\ref{eq:field_operator}) in Eq. (\ref{eq:Hee_space}), 
the electron-electron interaction in the momentum representation reads:
\begin{eqnarray}
H_{e-e}&=&\frac{1}{8A}
\sum_{\{\lambda_i\}}\sum_{\bm k,\bm k',\bm q}
\sum_{\sigma,\sigma'}
\varphi_{\rm 2D}(\bm q) 
a^\dag_{\lambda_1,\bm k +\bm q,\sigma}
a^\dag_{\lambda_2,\bm k'-\bm q,\sigma'}\times\nonumber\\
&&a_{\lambda_3,\bm k',\sigma'}
a_{\lambda_4,\bm k,\sigma}
f(\bm k',-\bm q)_{\lambda_2,\lambda_3}
f(\bm k,\bm q)_{\lambda_1,\lambda_4}\,,
\label{eq:He-efinal}
\end{eqnarray}
 where
$
f(\bm k,\bm q)_{\lambda_a,\lambda_b}= 
(1+\lambda_a\lambda_be^{-i[\theta(\bm k+\bm q)-\theta(\bm k)]})\,.
$
As before, we seek the equation of motion for the 
operator $\hat P_{vc,\bm k}=a^\dag_{v,\bm k}a_{c,\bm k}$, but 
now in the presence of electron-electron interactions. To that end, 
we need to compute the commutator of $\hat P_{vc,\bm k}$ with $H_{e-e}$, 
in addition to the one we have already determined for the non-interacting theory. 
The evaluation of $\langle [H_{e-e},\hat P_{vc,\bm k}]\rangle$, after
making the usual Hartree-Fock decoupling the quartic terms, gives:
\begin{eqnarray}
\langle[H_{e-e},\hat P_{vc,\bm k}]\rangle\!\!&=&\!\!C^{\rm exch.}\!+
\! C^{\rm excit.}\!+\!C^{\rm dens.}+
C^{\rm n-l}\,,
\label{eq:commutator}
\end{eqnarray}
where the several terms have the following physical 
interpretation: $C^{\rm exch.}$ is the exchange correction to 
the bare bands; $C^{\rm excit.}$  are the 
excitonic contributions, coming from the Coulomb interaction among  the 
electron-hole pair formed by the absorption of light; $C^{\rm dens.}$ is 
the interaction of the electrons and holes in the bands with the electron gas
 formed by the gating; and  $C^{\rm n-l}$ are
 non-linear terms in the field intensity, $E_0$, due to the interaction of the average 
of the operator  $\hat P_{vc,\bm k}$ with itself and with the average of the 
operator $\hat P_{cv,\bm k}$ (see also Ref. \onlinecite{Mishchenko}).

Let us consider the calculation of the vertex corrections, 
$\Gamma_{\lambda\lambda',\bm k}$, entering in Eq. (\ref{eq:sigmaxx}). We ignore the
 non-linear corrections included in the commutator (\ref{eq:commutator}). The 
term $C^{\rm density}$ can be shown to give a zero contribution to the polarization
 of the material. From the commutator (\ref{eq:commutator}), the full equation of
 motion for the operator $\hat P_{vc,\bm k}$ gives:
\begin{eqnarray}
 &&(-\omega\hbar-E_{v,\bm k}+E_{c,\bm k})\langle\hat P_{vc,\bm k} \rangle=
E_0\frac{v_Fe}{i\omega}d^x_{cv,\bm k}\delta n_{F,\bm k}
\nonumber\\
&+&\delta n_{F,\bm k}\frac{1}{2A}
\sum_{\bm q,\sigma}
\varphi_{\rm 2D}(\bm q)\langle\hat P_{vc,\bm k-\bm q}\rangle
F_{+,+} (\bm k,\bm q)
\nonumber\\
&+&
\delta n_{F,\bm k}\frac{1}{2A}
\sum_{\bm q,\sigma}
\varphi_{\rm 2D}(\bm q)\langle\hat P_{cv,\bm k-\bm q}\rangle
F_{-,-} (\bm k,\bm q)\,,
\label{eq:eq_motion_polarization}
\end{eqnarray}
with $\delta n_{F,\bm k}=[n_F(E_{c,\bm k})-n_F(E_{v,\bm k})]$
 and $F_{\lambda,\lambda}(\bm k,\bm q)=1+\lambda\cos[\theta(\bm k)-\theta(\bm k-\bm q)]$. 
Eq. (\ref{eq:eq_motion_polarization}) and the equivalent one for 
$\langle\hat P_{cv,\bm k}\rangle$, obtained from Eq. (\ref{eq:eq_motion_polarization}) 
by interchanging the labels $c$ and $v$, form a set of two coupled integral equations. 
The approximate solution of this set of equation is obtained as follows: since we are 
interested in the linear response, we write $ P_{vc,\bm k}=E_0\chi_{vc,\bm k}$. Additionally, 
the function $\chi_{vc,\bm k}$ is written in terms of the $\chi^0_{vc,\bm k}$ 
(which represents $\chi_{vc,\bm k}$ in the absence of electron-electron interactions) 
as $\chi_{vc,\bm k}=\Gamma_{vc,\bm k}\chi^0_{vc,\bm k}$, where $\Gamma_{vc,\bm k}$ is 
the vertex function. After these operations, a set of coupled integral 
equations for the vertex functions is obtained from Eq. (\ref{eq:eq_motion_polarization}), 
 which is then solved by the method of Pad\'e approximants \cite{Haug1}. For simplicity, we 
have only considered the leading approximant. The method of solution described here is 
equivalent to summing an infinite sub-set of Feynman diagrams, 
and it gives the analytical solution for the vertex in the 
form $\Gamma_{\lambda \lambda',\bm k}\approx[1+s_{vc,\bm k,1}]^{-1}$ with 
the $s_{\lambda \lambda',\bm k,1}$ defined as
\begin{eqnarray}
s_{\lambda \lambda',\bm k,1} \!\!&=&\!\! -\! \frac{\omega}
{v_Fe\sin\theta_{\bm k}}\Theta_{\lambda \lambda',\bm k}^+
\!\!-\!\!\frac{\omega}{v_Fe\sin\theta_{\bm k}}\Theta_{\lambda' \lambda,\bm k}^- \,.
\label{eq:scvk}
\end{eqnarray}
where, 
\begin{equation}
\Theta_{\lambda \lambda',\bm k}^{\pm} \!\!=\!\!\frac{1}
{2\pi^2}\int_{0}^{2\pi}\!\!d\theta_{\bm q}
\int_0^{q_c}\!\!qdq
\varphi_{\rm 2D}(\bm q-\bm k)\chi^0_{\lambda \lambda',\bm q}[1\pm \cos(\theta_{\bm k}\!-\!\theta_{\bm q})] \, . 
\label{eq:Thetavc}
\end{equation}
This procedure for determining the vertex function is equivalent to the
RPA solution of the 
Bethe-Salpeter equation \cite{hanke}; we believe, however, that our method
 is more straightforward.
 Equations (\ref{eq:scvk}) and (\ref{eq:Thetavc}) can be used to
 determine the vertex function in closed form. The bare 
susceptibility, $\chi^0_{\lambda\lambda',\bm k}$, 
is given by
\begin{eqnarray}
\chi^0_{\lambda \lambda',\bm k}\!=\!-\sin\theta_{\bm k}\frac{ev_F}{\omega}\!
\left(\! \frac{n_F(E_{\lambda,\bm k})\!-\!n_F(E_{\lambda',\bm k})}
{\hbar\omega\!-\!E_{\lambda',\bm k}\!+\!
E_{\lambda,\bm k}\!+\!i\Gamma} \!+\! \lambda' \frac{\theta(k)}{2v_F\hbar k}
\right) \, ,
\end{eqnarray}
where $\theta(k)$ is the Heaviside step function.
Unfortunately the integrals cannot be evaluated analytically but 
can easily be obtained by numerical integration.

In summary, we have developed a general formalism for taking into account 
excitonic effects in graphene. These effects were computed taking the lowest Pad\'e 
approximant, but higher order ones can be easily introduced
and may improove the fit of the data. Additional many-body effects, 
such as inelastic phonon scattering can also be easily included, which makes the formalism a 
powerful one to study the combined effects of disorder, electronic correlations, and 
phonon scattering in graphene's optical spectrum. Extending the formalism to visible 
range of the spectrum is also feasible, by computing the eigenproblem of the free 
theory including trigonal warping effects. Finally, we have showed that our  formalism 
solves discrepancies between the data \cite{basov} and the theoretical results 
provided by the independent electron picture.

AHCN acknowledges DOE grant DE-FG02-08ER46512 and ONR grant MURI N00014-09-1-1063.
NMRP acknowledges the FCT grant PTDC/FIS/111524/2009, and T. Stauber and 
M.  Vasilevskiy for comments at an earlier stage of the manuscript.

\bibliographystyle{apsrev} 
%

\end{document}